\documentclass[]{interact}
\usepackage{array}
\usepackage{amsmath}
\usepackage{tikz}
\usepackage{pgfplots}
\usepackage{tabularx}
\usepackage{textcomp}
\usepackage{comment}
\usepackage{csquotes}
\usepackage[hidelinks]{hyperref}
\usepackage{booktabs}
\usepackage{appendix}
\usepackage[longnamesfirst,sort]{natbib}
\bibpunct[, ]{(}{)}{;}{a}{,}{,}

\usetikzlibrary{arrows.meta,fit,positioning}
\pgfplotsset{compat=1.18}
\newcommand{\EXCLUDE}[1]{}

\newcommand{\Xonetwothree}{X1-X2-X3} 
\newcommand{\Xott}{\Xonetwothree}

\begin{document}
\title{%
Assessment Design in the GenAI Era: %
The X1-X2-X3 Assessment Pattern for Testing Students' AI Literacy, Learning Outcomes, and Reflection
}

\author{
\name{Riasat Islam\textsuperscript{a}\thanks{Corresponding author. Email: \href{mailto:riasat.islam@qmul.ac.uk}{riasat.islam@qmul.ac.uk}. ORCID: \href{https://orcid.org/0000-0002-1419-8068}{0000-0002-1419-8068}.} and Thomas Roelleke\textsuperscript{a}\thanks{ORCID: \href{https://orcid.org/0009-0009-8879-6587}{0009-0009-8879-6587}.}}
\affil{\textsuperscript{a}School of Electronic Engineering and Computer Science, Queen Mary University of London, London, United Kingdom}
}

\maketitle

\begin{abstract}
Generative artificial intelligence (GenAI) has challenged the validity of unsupervised online assessment, especially in technical subjects where plausible answers can be produced with little effort. This paper reports lessons from designing and implementing an AI-aware, AI-testing assessment in a large second-year undergraduate database systems module. The design combined two linked elements: (1) a structured three-part response format (X1-X2-X3) in which students documented a sourced answer, produced their own answer, and evaluated the sourced output; and (2) an AI-aware question-design process in which draft tasks were stress-tested against contemporary GenAI tools and revised when generic prompting produced superficially adequate answers. The account draws on archived assessment materials, rubrics, planning records, design-time GenAI trials, practice-response data, attainment records, and external review comments. Its main contribution is a reusable assessment-design method rather than a claim of measured learning gains. We show how the pattern developed across iterations and how it can support authentic assessment, visible AI literacy, student judgement, and more transparent marking. The paper offers practical guidance for lecturers adapting assessment to routine GenAI use, focusing on testing AI literacy rather than penalising students for misconduct.
\end{abstract}


\section{Introduction}
Higher education assessment is experiencing a capability shift. Tasks that once served as plausible proxies for individual understanding can now often be completed to a high surface standard using widely available GenAI tools. The resulting problem is not simply that students can access AI; it is that many existing assessments no longer make the intended constructs of learning sufficiently visible. In the United Kingdom, this challenge is sharpened by rapid normalization of student AI use, continuing post pandemic concern about the interpretability of assessment outcomes, and the persistence of online conditions in which covert assistance is difficult to observe directly. \citet{hepi2025} and \citet{hepi2026} report that student AI use has normalized with exceptional speed: 92\% of undergraduates used AI in some form in 2025 and 95\% did so in 2026, while use of GenAI to help with assessed work rose from 88\% to 94\% across the same period. \citet{ofs2024,ofs2026} likewise document large post pandemic changes in degree classifications in England, while \citet{newton2024} and \citet{lancaster2021} report increased cheating pressure in online assessment environments. We use those sector level trends as context rather than as outcomes of the present case study. Their relevance here is narrower: they intensify the need for assessment designs whose evidential basis remains interpretable when routine AI use is part of the background conditions.

The emerging literature no longer treats this as a niche concern. policy oriented and review work consistently argues that GenAI requires redesign of assessment rather than reliance on prohibition alone \cite{chan2023,xia2024,luo2024}. Sector guidance similarly emphasizes human agency, ethical use, AI literacy, and institutional responsibility to adapt assessment in ways that preserve rigour while acknowledging the reality of tool availability \cite{unesco2023,webb2023,walker2025,qaa2022computing,russell2023}. In computing education, that challenge is especially acute. Database systems is a revealing site because it spans several task types at once: procedural SQL, conceptual data modeling, normalization, and broader architectural reasoning. GenAI can often generate fluent initial answers for each of these forms, yet the technically important distinctions frequently lie in omitted assumptions, wrong abstractions, or weak conceptual decomposition. What remains educationally valuable, therefore, is not merely whether an answer can be produced, but whether it can be evaluated against disciplinary expectations.

Response format, however, is only part of the validity problem. Under current AI capability, question design itself becomes a primary assessment mechanism. Generic textbook style questions can often be completed by LLMs at a superficially acceptable level through pattern matching over familiar forms, even when deeper disciplinary reasoning is weak. If a task lacks contextual specificity, local assumptions, or subtle discriminators, it may invite answer retrieval rather than the intended performance. In this case study, we therefore treated question authoring as an AI-aware design activity as well as a marking problem: members of the module team tested draft questions with contemporary GenAI tools, examined how easily plausible answers could be generated, and revised questions when superficial completion appeared too easy.

In this paper, we address that challenge through a case from an undergraduate database systems module (270 students) at a UK research-intensive university. Institutional and personnel identifiers are withheld for anonymous review. Rather than banning web search and GenAI outright, we developed and refined an AI-integrated assessment design in which students had to (1) provide a sourced response and respective evidence, (2) produce their own answer and show their understanding of LOs, and (3) reflect on the process and assess the sourced answer. We stabilized this as a three-part response structure, hereafter \texttt{\Xott}, or, short, X123. We iterated it across four deployments: coursework practice, assessed coursework, practice exam, and final exam.
This iteration was essential since the exam format was very new to students.
Conversation with students included:
\begin{quote}
    Student: Really, we can use AI in the exam? You mean it is not like before (where we used it but it was not allowed), but now it is allowed?
    [Yes, not just allowed; it is requested.]

    [After a practice session]: Student: Well, it is really great that we can use it. But you are assessing how we use it! Do we have to know if there is something wrong with the AI answer? [Yes, that is part of it. You must demonstrate the LOs of the module.]
\end{quote}

We position this case study as iterative practice based assessment design research. Our aim is not to claim causal effects from a controlled intervention, but to articulate a theoretically grounded and operationally tested assessment pattern that other educators can adapt in comparably complex, real teaching environments \cite{dbrcollective2003,wang2005}. The paper should therefore be read primarily as an assessment design contribution in computing education, with one database systems module serving as the empirical site through which the design was developed and scrutinized. We address the following research questions:
\begin{enumerate}
\item \textbf{RQ1:} How were question and response format refined across four deployments into a scalable AI-aware three-part assessment structure (\texttt{\Xott})?
\item \textbf{RQ2:} How did the resulting combination of AI-aware questions and the three-part response structure (\texttt{\Xott}) make AI use visible, reviewable, and markable in ways aligned with evaluative judgement and authentic assessment?
\item \textbf{RQ3:} What do the documentary evidence and descriptive outcome data suggest about attainment patterns, workload, and design trade-offs?
\end{enumerate}
We make three contributions. First, we present a reusable assessment design framework that combines AI-aware question authoring with a three-part response structure (\texttt{\Xott}). Second, we formalize a practical authoring method in which candidate tasks are stress-tested against contemporary GenAI tools and revised when superficial prompting yields overly adequate answers. Third, we report descriptive design lessons from iterative deployment at scale, including what became visible in student responses, where workload shifted, and which implementation details mattered most for markability, fairness, and technical specificity.

\section{Related Work and Theoretical Foundation}
\subsection{Generative AI, Integrity, and Assessment Redesign}
Recent scholarship on GenAI in higher education has converged on a broad conclusion: assessment validity can no longer be protected by assuming that students will not use AI, or that simple detection will solve the problem. \citet{chan2023} argues for institutionally coherent AI policy that addresses pedagogy, governance, and assessment together. \citet{luo2024} similarly shows that higher education is still renegotiating what originality, authorship, and acceptable assistance mean in the presence of LLMs. \citet{xia2024} extend this argument by showing that GenAI is reshaping assessment across student, teacher, and institutional levels, making redesign an urgent rather than optional task.

Academic-integrity commentary reaches similar conclusions. \citet{kirwan2024} argues that GenAI forces universities to reconsider what assessment is actually measuring. \citet{cotton2024} likewise show that ChatGPT-like tools intensify familiar integrity risks while rendering blanket bans pedagogically weak and operationally brittle. Taken together, this literature suggests that the main problem is not merely misconduct; it is that product only tasks increasingly fail to reveal whether students themselves exercised understanding and judgement.

\subsection{Rethinking Assessment Design: Mandatory AI Usage Instead Of Detecting AI Offences}

Taken together, this literature suggests that assessment redesign is a more defensible response than detector led attribution. \citet{webb2023,webb2025detect} repeatedly cautions that AI-detection tools cannot conclusively prove authorship and should not be treated as a dependable basis for academic judgement. That caution is reinforced by empirical work showing that detector performance is not only imperfect but unevenly distributed: \citet{liang2023} found that widely used GPT detectors were strongly biased against non-native English writing. In practice, the educational problem is therefore poorly served by a model of assessment that assumes authorship can be inferred after submission. A more defensible alternative is to make tool use visible inside the task itself, so that what is assessed is not hidden provenance but the student's ability to interpret, revise, and justify an answer under transparent conditions.

\subsection{Authentic Assessment and Constructive Alignment}
Authentic assessment provides one productive response to that problem. \citet{villarroel2018} describe authentic assessment as assessment that meaningfully reflects the practices and standards of professional contexts. \citet{sotiriadou2020} connect authentic assessment to employability and academic integrity, arguing that carefully designed authentic tasks can both preserve standards and develop valuable graduate capabilities. At the same time, \citet{ellis2020} warn that authenticity alone does not guarantee integrity; authentic tasks must still make the student's own contribution and judgement visible. Recent work on digital authenticity reinforces this point, arguing that authenticity in digitally mediated assessment must be designed rather than assumed \cite{nieminen2023,ajjawi2024}.

These arguments align naturally with constructive alignment. \citet{biggs1996} argues that assessment should be aligned with intended learning outcomes and teaching activities, so that the task requires the student to demonstrate the forms of understanding the course claims to value. In an AI-rich environment, many conventional online tasks become misaligned if they can be completed persuasively without the student demonstrating the target reasoning. An AI-integrated assessment pattern is therefore only valid if it re-specifies the target performance so that critical judgement, not just output production, becomes visible. This is especially relevant where benchmark guidance expects graduates to locate and retrieve relevant ideas, communicate them effectively, reflect critically, and engage with authentic assessment opportunities connected to professional practice \cite{qaa2022computing}. Recent sector frameworks such as the Artificial Intelligence Assessment Scale and the QAA's sustainable-assessment guidance point in the same direction by recommending that AI use be designed into assessment explicitly rather than treated as an afterthought \cite{perkins2024,qaa2023reconsidering}.

\subsection{Question Design Under AI Capability Shift}
The same logic applies to question design itself. Recent GenAI-era guidance repeatedly recommends more authentic, higher-order, and context rich assessment, especially where generic prompts are vulnerable to direct completion \cite{perkins2024,qaa2023reconsidering,ajjawi2024}. However, much of this literature remains prescriptive at a high level: it argues that assessment must change, but says less about an operational process by which educators can test whether a draft question is already too easy for current GenAI systems. From a constructive-alignment perspective, this is a genuine gap. If a draft task can be answered plausibly through superficial prompting, then the validity problem begins before student response format is considered.

Our contribution here is therefore not the claim that scenario based or contextualized questions are new. Rather, it is a practical AI-aware question design method for the GenAI era: educators use contemporary GenAI tools as a baseline during assessment authoring, inspect where answers are deceptively plausible or almost complete, and revise prompts toward context rich, module specific formulations that require interpretation, discrimination, and judgement. This operationalizes the broader literature on authenticity, alignment, and evaluative judgement in a way that is directly usable by teaching teams.

\subsection{Evaluative Judgement, Self-Regulated Learning, and Graduate Attributes}
Our theoretical framing draws on critical-thinking scholarship, but the more directly operational construct in this paper is evaluative judgement. \citet{facione1990}'s Delphi report is helpful background because it defines critical thinking as purposeful, self-regulatory judgement involving interpretation, analysis, evaluation, inference, and explanation. However, our evidence does not support a claim that we measured or improved critical thinking as a broad trait. What we can examine more directly is whether the assessment made evaluative comparison, revision, and justification visible.

\citet{tai2018} define evaluative judgement as the capability to make decisions about the quality of work. Recent work extends this directly into the GenAI era: \citet{bearman2024} argue that learners must judge not only AI outputs but also the processes used to obtain them, including the credibility of prompts, iterations, and comparisons against disciplinary standards. \citet{zimmerman2002} frames self-regulated learners as active agents who plan, monitor, and evaluate their own learning processes. The reflective component of our design also draws on work that treats reflection as the process by which experience is revisited and turned into learning \cite{boud1985}. \citet{barrie2007}, in turn, shows that graduate attributes such as communication, ethical reasoning, and critical scrutiny only become meaningful when they are embedded in disciplinary curriculum and assessment rather than left as abstract aspirations. Our assessment design therefore treats AI literacy and evaluative judgement not as extra-policy aspirations, but as capabilities to be enacted within technical-subject assessment itself.

Taken together, these literatures point to a strong design principle for AI-integrated assessment: students should not receive credit merely for obtaining a plausible answer from a tool. They should receive credit for making tool use transparent, revising the output through disciplinary reasoning, and demonstrating evaluative judgement about what the tool produced.

\section{Case Study and Context}

\subsection{Case Study: Database Exam, 2025/26}

We treat this case study as educational research in the sense that we iteratively designed a live assessment intervention while also attempting to extract transferable assessment design knowledge from that process \cite{dbrcollective2003,wang2005}. The case study is therefore best understood as practice based assessment design research rather than as an experiment. RQ1 focuses on evolution across the four deployments: coursework practice, assessed coursework, practice exam, and final exam. RQ2 focuses on the operational mechanism by which the final assessment design rendered students' use of web and GenAI tools visible and assessable; and RQ3 focuses on the descriptive consequences and trade-offs that emerged once the assessment was used at scale.

\subsection{Module Context and Assessment Design Challenge}
We studied a large undergraduate database systems module at a UK research-intensive university. For the run reported here, the official module-result export contained 275 students, of which 263 had complete component marks for the descriptive analyses reported here; detailed question level coursework analysis was available for 251 students. The module included online coursework, practice tasks, a practice exam, and a final online exam. This was therefore a realistic, high-volume teaching context rather than a small pilot.

The broader student context is also relevant. The module was taught in English at an English-medium UK university and enrolled undergraduate students from a culturally and ethnically diverse cohort. Routine admissions records indicated substantial representation of both home and international students. We report that background only as contextual information, however: the present case study did not undertake a dedicated demographic analysis or subgroup comparison by domicile, ethnicity, or language background.

The challenge was to preserve the validity of online assessment in a context where students could readily consult the internet or GenAI tools in ways that were difficult to observe and regulate. A blanket prohibition would neither reflect actual student practice nor align well with current sector guidance on AI literacy, ethical use, and equitable access \cite{unesco2023,russell2023,hepi2026}. The central question was therefore not whether GenAI existed, but how assessment could be designed so that students' own disciplinary reasoning and judgement remained the primary object of evaluation. In practice, this required treating web and GenAI outputs as provisional resources to be interrogated, adapted, and evaluated by students rather than as authoritative answers to be reproduced.

\subsection{Assessment Design Goals}
We pursued six design goals.
\begin{enumerate}
\item \textbf{Validity under AI capability:} the assessment still had to measure database reasoning rather than mere acquisition of a plausible answer, consistent with constructive alignment \cite{biggs1996}.
\item \textbf{Question robustness under contemporary GenAI:} draft questions needed to be less susceptible to superficial GenAI completion and better able to elicit interpretation rather than direct retrieval.
\item \textbf{Visibility of evaluative judgement:} the task needed to elicit evaluation, explanation, and judgement, not just answer production \cite{facione1990,tai2018}.
\item \textbf{Authenticity and graduate attributes:} the format should mirror realistic professional problem solving, where engineers consult online sources and increasingly use AI tools, but remain accountable for correctness and trade-offs \cite{villarroel2018,sotiriadou2020,barrie2004,qaa2022computing}.
\item \textbf{Transparency and reviewability:} the assessment should make tool use inspectable through source evidence and explanation rather than relying primarily on detector-style attribution \cite{chan2023,luo2024,liang2023,webb2025detect,russell2023}.
\item \textbf{Scalability and equity:} the design had to work for a large cohort with manageable marking load and without advantaging students simply because they had access to better paid tools \cite{hepi2026,unesco2023}.
\end{enumerate}

These goals were translated into a smaller set of design principles that connected the theoretical framing, the module-level implementation, and the intended assessment benefit. Table~\ref{tab:principles} summarises this link between theory and the implemented assessment design.

\begin{table*}[t]
\caption{Design principles linking theory to AI-aware question design and the \Xott{} response structure}
\label{tab:principles}
\centering
\footnotesize
\begin{tabularx}{\textwidth}{p{2.4cm} p{3.2cm} >{\raggedright\arraybackslash}X >{\raggedright\arraybackslash}X}
\toprule
Design principle & Theoretical basis & Operationalization in the module & Intended assessment benefit \\
\midrule
AI-aware question design & Constructive alignment, authentic assessment, digital authenticity \cite{biggs1996,villarroel2018,ajjawi2024} & Draft questions were stress-tested against contemporary GenAI tools and revised when generic or textbook-like prompts yielded plausible almost complete answers; revisions added scenario context, module specific schemas, and subtle discriminators. & Reduces susceptibility to superficial GenAI completion and preserves the need for interpretation, trade-off awareness, and disciplinary judgement. \\
Transparency over detection & Integrity-by-design, detector limitations, evaluative judgement \cite{liang2023,webb2025detect,tai2018} & Students had to show prompts, outputs, source trace, timestamp, and a readable full-screen screenshot in \texttt{X1}. & Makes AI use auditable and discussable without asking markers to infer hidden authorship after submission. \\
Judgement over retrieval & Evaluative judgement and critical scrutiny \cite{facione1990,tai2018,bearman2024} & \texttt{X2} required a student authored revision; \texttt{X3} required a bounded rating and justification of the sourced answer. & Rewards diagnosis, revision, and justification rather than fluent copying of plausible text. \\
Authenticity with disciplinary specificity & Authentic assessment, constructive alignment, benchmark \cite{villarroel2018,biggs1996,qaa2022computing} & Questions were framed around technical database scenarios where GenAI could help but full marks still depended on precise disciplinary distinctions, interpretation of constraints, and use of module specific concepts. & Preserves real-world relevance while keeping the target learning outcomes technically specific. \\
Scalability with fairness & Graduate attributes, sustainable assessment, equitable access \cite{barrie2007,qaa2023reconsidering,russell2023} & Standard \texttt{X1/X2/X3} structure, word limits, common screenshot rules, reusable rubric, and expectation of free tier tool use. & Reduces verbosity, supports consistent marking, and limits advantage from premium subscriptions. \\
\bottomrule
\end{tabularx}
\end{table*}

\subsection{AI-aware Question Design through Stress Testing}
Alongside designing the \Xott{} response format, we designed the questions themselves through an AI-aware question design process. Internal planning documents and stored design-time GenAI answer files show that members of the module team routinely attempted exam questions with contemporary GenAI tools before deployment. In doing so, we used GenAI tools to test whether draft questions could be answered too easily through generic, superficially plausible responses. The purpose was not to make questions impossible for AI systems, nor to ban AI use in principle. Rather, it was to identify when a draft task was so generic that GenAI could produce an apparently acceptable answer with little effort, thereby weakening the validity argument for the assessment.

In practice, the authoring cycle had five recurring steps. First, we drafted exam questions aligned to the learning outcomes, for example, \enquote{list and explain the data warehouse schemas}. Second, members of the module team attempted those questions with the kinds of GenAI tools students were likely to use, for example ChatGPT, Gemini, Claude, and DeepSeek. Third, we inspected the resulting answers for correctness, completeness, and deceptively plausible omissions. Fourth, we revised the tasks when GenAI could answer them too readily through generic pattern matching rather than genuine disciplinary reasoning. For example, this led to rephrasing the data warehouse question to \enquote{list and classify the main data warehouse schemas}. Fifth, we re-tested the revised questions before final deployment. This cycle ran alongside the refinement of the marking scheme and the \Xott{} response structure. Table~\ref{tab:authoringcycle} summarizes the resulting protocol in a more reproducible form.

\begin{table}[h!tbp]
\caption{AI-aware question authoring cycle used during assessment design}
\label{tab:authoringcycle}
\centering
\scriptsize
\begin{tabularx}{\columnwidth}{@{}p{1.0cm} >{\raggedright\arraybackslash}X >{\raggedright\arraybackslash}X@{}}
\toprule
Step & Authoring question & Typical design move \\
\midrule
1. Draft & What disciplinary performance should this task elicit, and what would a superficial sourced answer look like? & Draft the question against the learning outcome and identify the likely role of web/GenAI support. \\
2. Test & What happens if the task is attempted with currently available GenAI tools using generic prompting? & Run one or more tool trials using the kinds of prompts students are likely to try first. \\
3. Inspect & Is the answer correct, almost complete, deceptively plausible, or weak in technically important ways? & Check for generic pattern matching, omitted discriminators, shallow decomposition, or fluent but incomplete reasoning. \\
4. Trigger & Is the task too easy to complete superficially, or too weakly specified to require judgement? & Add scenario context, locally authored schema elements, module specific terminology, edge cases, trade-offs, or assumptions that must be interpreted. \\
5. Re-test & Does the revised task test the respective learning outcomes while making superficial completion less adequate? & Re-run a tool trial and revise again if needed before deployment. \\
\bottomrule
\end{tabularx}
\end{table}

Several signals triggered redesign. Exam questions were revised when GenAI produced a almost complete answer from a generic prompt, when the answer looked fluent but relied on stock textbook patterns rather than interpretation, when the question lacked sufficient contextual specificity, or when the task did not require judgement about assumptions, trade-offs, or subtle distinctions. In response, we moved toward more scenario based and context rich prompts, embedded module specific schemas, terminology, and constraints, and introduced small discriminators or edge cases that required careful reading rather than retrieval alone. In database questions, this often meant using locally authored schemas, short professional scenarios, or technically meaningful distinctions that were not readily resolved by generic recall.

scenario based and context rich questions mattered because they altered the role of GenAI. When a prompt depended on local context, embedded assumptions, or module specific discriminators, GenAI could still provide a useful starting point, but it became less reliable as a direct answer engine. The task shifted from answer extraction toward interpretation and judgement. This complemented the three-part response structure directly: AI-aware question design reduced the likelihood that a student could succeed through superficial prompting alone, while \texttt{\Xott} made any subsequent tool use visible and assessable.

\subsection{The \Xott{} Response Structure}
In the early coursework version of the assessment, we asked students first to obtain an initial answer or approach from a web or GenAI source; then to document the source with evidence; then to produce their own answer; then to explain the difference between their answer and the sourced answer; and finally to reflect on how the tool helped or misled them. As the assessment moved from coursework to exam use, we standardized this logic into the \texttt{\Xott} structure.

\begin{itemize}
\item \textbf{X1 (Extract (X1a) and Evidence (X1b)):} a concise extract from the sourced answer, source identification, timestamp, and a mandatory full-screen screenshot.
\item \textbf{X2 (Student's Answer (X2a) and Difference (X2b)):} the student's answer (this can be based on X1, of course) plus a concise account of how it is different from X1. If there is no difference, then the student must explain (in their own words) why X1 is perfect.
\item \textbf{X3 (Narrative (X3a) and Assessment (X3b)):} a brief reflection on what was learned, and a rating of the sourced answer on a familiar scale (e.g. 5* scale), including a justification.
\end{itemize}

The practice exam rubric set explicit word limits of 100 words for \texttt{X1a}, 200 words for \texttt{X2}, and 100 words for \texttt{X3}. The exam also standardized attachment expectations: the screenshot had to capture the entire screen, show the current question, and remain readable for marking. In this deployment, the exam used an 18-point rubric mapped across the three fields and later converted into question marks, although the precise point allocation can be adapted by question setters. Fig.~\ref{fig:x123} shows the underlying logic of the pattern.

\begin{figure}[h!tbp]
\centering
\begin{tikzpicture}[
box/.style={draw, rounded corners, thick, fill=black!5, align=left, text width=0.22\textwidth, minimum height=3.2cm, inner sep=8pt},
arrow/.style={-{Latex[length=3mm]}, thick},
]
\node[box] (x1) {\textbf{X1: Extract\\and Evidence}\newline
- 100 words extract from web/GenAI source\newline
- source and timestamp\newline
- screenshot according to format requested};
\node[box, right=0.85cm of x1] (x2) {\textbf{X2: Student's answer and Difference}\newline
- student's answer (can be a revision of X1); in \enquote{own words}\newline
- discipline-specific reasoning (demonstration of learning outcomes)\newline
- explanation of key differences};
\node[box, right=0.85cm of x2] (x3) {\textbf{X3: Narrative and\\Assessment}\newline
- what was learned\newline
- critique of the sourced answer\newline
- bounded quality rating (e.g. 5*) and rationale};

\draw[arrow] (x1.east) -- (x2.west);
\draw[arrow] (x2.east) -- (x3.west);

\node[draw, dashed, rounded corners, fit=(x1)(x2)(x3), inner sep=10pt,
label=below:{\textit{Process: transparent use of AI, disciplinary reasoning (LOs), evaluative judgement}}] {};
\end{tikzpicture}
\caption{The \Xott{} response structure: Student's moving from AI (X1) to their answer (X2), and then evaluative judgement/assessment (X3).}
\label{fig:x123}
\end{figure}

Table~\ref{tab:rubric} translates the \texttt{\Xott} structure into the operational answer template and scoring logic used in the stabilized version of the assessment.

\begin{table*}[h!tbp]
\caption{Operational answer template and scoring logic for the stabilized \Xott{} response structure}
\label{tab:rubric}
\centering
\footnotesize
\begin{tabularx}{\textwidth}{p{1.4cm} >{\raggedright\arraybackslash}X p{3.2cm} p{1.9cm}}
\toprule
Component & Student task & Constraint / evidence requirement & Points \\
\midrule
X1a & Paste the most relevant part(s) of the sourced answer and identify the service used. & Extract limited to 100 words; source and timestamp required. & 0--3 \\
X1b & Document the source in a way that a third party could inspect. & Full-screen readable screenshot showing the sourced answer, current question, and student response fields. & 0--3 \\
X2a & Produce a revised answer in the student's own terms. & Main answer field; expected to demonstrate disciplinary understanding rather than reuse alone. & 0--3 \\
X2b & Explain the main differences from X1, or justify why no difference was necessary. & Concise explanation, typically up to three main differences, within the 200-word X2 field. & 0--3 \\
X3a & Describe what was learned and assess the quality of X1. & Concise narrative. & 0--3 \\
X3b & Provide a bounded quality rating (e.g. 5*) and a short rationale. & & 0--3 \\
\bottomrule
\end{tabularx}
\end{table*}

As shown in Table~\ref{tab:rubric}, the stabilized rubric made the final assessment more standardized and reviewable than the earlier coursework version in two senses. First, students received a fixed response template with explicit word limits and evidence requirements. Second, markers applied a common rubric logic in which the sourced extract, evidence trail, revision, explanation of differences, and reflective judgement were all explicitly visible in the response; in the exam deployment reported here, this was operationalized through an 18-point scheme. In both the practice and summative exam versions, the marking emphasis was on the student's judgement and documentation, not on the intrinsic superiority of the external tool they happened to use.

To illustrate how this scoring logic separated retrieval from disciplinary judgement, Table~\ref{tab:workedexample} gives a representative worked example based on the intended response moves across \texttt{X1}, \texttt{X2}, and \texttt{X3}.

\begin{table*}[h!tbp]
\caption{Representative worked example of how the three-part response structure (\texttt{\Xott}) distinguishes retrieval from disciplinary judgement}
\label{tab:workedexample}
\centering
\footnotesize
\begin{tabularx}{\textwidth}{p{1.2cm} >{\raggedright\arraybackslash}X >{\raggedright\arraybackslash}X}
\toprule
Field & Representative response move & What the marker could infer \\
\midrule
X1 & Student submits a sourced conceptual model that bundles \texttt{Source(Id, Owner, Provider)} into one entity and documents the prompt, source, and screenshot correctly. & The student can retrieve and document a superficially plausible answer, but the submission does not yet demonstrate conceptual discrimination. \\
X2 & Student revises the answer to separate \texttt{Source}, \texttt{Owner}, and \texttt{Provider} into distinct entities with explicit relationships and explains that the sourced answer conflated entities with attributes. & The student demonstrates disciplinary reasoning, revises the sourced answer rather than repeating it, and identifies the conceptual modeling issue that matters in database design. \\
X3 & Student rates the sourced answer low or mid-scale and justifies the rating by explaining that it is useful as a starting point but weak conceptually because it collapses distinct design objects. & The student makes an evaluative judgement about the adequacy of the sourced answer and articulates why the revision in X2 is superior. \\
\bottomrule
\end{tabularx}
\end{table*}

This worked example captures the logic we intended across the broader assessment set. Credit was not tied to whether the external tool sounded fluent, but to whether the student could diagnose what was technically weak, revise it, and justify the revision concisely.

\subsection{Iterative Refinement}
Table~\ref{tab:iterations} summarizes the iterative trajectory.

\begin{table*}[h!tbp]
\caption{Iterative refinement of the \Xott{} response structure across four assessment instances}
\label{tab:iterations}
\centering
\footnotesize
\begin{tabularx}{\textwidth}{p{1.8cm} p{2.2cm} >{\raggedright\arraybackslash}X >{\raggedright\arraybackslash}X}
\toprule
Iteration & Setting & Main design features & Main lesson carried into the next iteration \\
\midrule
1 & Coursework practice & Students sourced a web/GenAI answer, documented the source, wrote their own answer, explained differences, and reflected on the process; exam questions were still relatively generic. & Revealed that students needed rehearsal, that markers needed tighter expectations around evidence and answer length, and that some draft questions remained too easy for superficial GenAI completion. \\
2 & Assessed coursework & The same core logic was applied to SQL and database design essay questions inside the module quiz, with early movement toward more context rich and module specific prompts after internal GenAI testing. & Showed that long answers, varied attachment formats, weak reflection, and overly generic prompts created marking burden and let some students rely too heavily on copied or weakly revised material. \\
3 & Practice exam & The response format was standardized into \texttt{X1/X2/X3}, with word limits, mandatory screenshot rules, explicit judgement of the sourced answer, and further stress-testing of exam questions against GenAI. & Improved clarity for students, reduced verbosity, and showed that scenario based and context rich prompts worked better when paired with a reusable response template. \\
4 & Final exam & The refined format was embedded into a mixed traditional/Web-GenAI online exam, alongside explicit instructions, structured marking, fairness constraints on tool access, and a question set revised through AI-aware authoring. & Produced the most stable version of the combined design and clarified which parts of the response structure and question design most directly supported rigour, efficiency, and critical engagement. \\
\bottomrule
\end{tabularx}
\end{table*}

We frame this case study as practice based assessment design education research rather than a controlled experiment: we aimed to solve a live pedagogic problem while refining a reusable assessment pattern in situ \cite{dbrcollective2003,wang2005}. We gathered feedback from students, the teaching team, teaching assistants and demonstrators, colleagues, and external review. Internal exam documentation records that external feedback commended the use of web/GenAI questions while also observing that conventional product only items remained less transparent about students' process.

We therefore present the resulting \Xott{} framework as a practice based, iterative refinement. Early iterations established the core pedagogic idea; later iterations improved both the operational fit of the three-part response structure and the robustness of the question set under contemporary GenAI capability.

\subsection{Evidence Base and Analysis}
We draw on seven forms of evidence:
\begin{enumerate}
\item contemporaneous design notes prepared during paper planning;
\item coursework and exam assessment artifacts, including question documents, instructions, and rubrics;
\item internal design documents such as the generic marking scheme and final exam planning materials;
\item stored sample GenAI answers and question-testing artifacts used while designing and validating questions;
\item an archived practice-response export containing student \texttt{X1/X2/X3} texts from the practice exam quiz;
\item aggregate coursework and module-result spreadsheets, including coursework final-grade exports and question level marks; and
\item external review comments on the exam design.
\end{enumerate}

These materials were not equally complete across the four iterations. Documentation is richest for the later practice exam and summative-exam phases, where rubrics, planning sheets, and marking documents survive systematically; the earliest coursework-practice phase is reconstructed more heavily from archived task wording, notes, and later reflections. Likewise, the stored design-time GenAI trials represent exemplars from each design phase rather than an exhaustive archive of every discarded draft question. We therefore use the corpus to support bounded descriptive claims, not to imply full coverage of every design decision.

Our analysis is descriptive rather than inferential. We use documentary evidence to reconstruct the design process, we use response traces to examine what kinds of action became visible within \texttt{X1/X2/X3}, and we use aggregate marks to check whether the resulting assessments showed an obvious ceiling pattern. We report pattern-level observations from design and marking, but we use the quantitative results to characterize variation rather than to claim causal proof.

Table~\ref{tab:corpus} summarises the evidence corpus, its coverage, its analytic role, and the main limitations associated with each source type.

\begin{table*}[h!tbp]
\caption{Evidence corpus, coverage, and analytic role}
\label{tab:corpus}
\centering
\footnotesize
\begin{tabularx}{\textwidth}{p{3.1cm} p{3.0cm} >{\raggedright\arraybackslash}X >{\raggedright\arraybackslash}X}
\toprule
Source type & Coverage in archive & Used mainly for & Main limitation \\
\midrule
Design notes and planning documents & Contemporaneous notes and planning sheets spanning all four iterations, with fuller documentation for the later exam phases & Reconstructing why questions, rubric expectations, and evidence rules changed over time & Earliest phases are less densely documented and require retrospective reconstruction \\
Assessment artifacts and rubrics & Coursework and exam question sheets, instructions, practice materials, and rubric documents for markers & Documenting the response format, question wording, and operational constraints actually deployed & Show intended design, not directly what students did with it \\
Stored design-time GenAI trials & Archived tool outputs captured during question testing across coursework and exam preparation & Showing what triggered redesign during AI-aware question authoring & Opportunistic rather than exhaustive archive of all authoring trials \\
Archived practice-response export & 25 student respondents, 50 attempted triads, 22 substantively complete triads & Describing completion patterns and visible response moves within \texttt{X1/X2/X3} & Practice exam quiz sample only; not a formal interview or think-aloud dataset \\
Coursework and module records & question level coursework marks ($n=251$), coursework exports ($n=260$), module records ($n=275$; 263 complete cases) & Characterizing attainment distributions and whether marks were heavily compressed at the top end & No pre/post or alternative-format comparator \\
External review comments & Exam-review feedback archived as part of routine quality processes & Providing an additional non-author commentary on the design and its practical consequences & Sparse and evaluative rather than systematic research data \\
\bottomrule
\end{tabularx}
\end{table*}

Table~\ref{tab:rqalignment} then maps each research question to the primary evidence sources, analytic use, and claims supported by the available data.

\begin{table*}[h!tpb]
\caption{Alignment of research questions, evidence sources, and analytic use}
\label{tab:rqalignment}
\centering
\footnotesize
\begin{tabularx}{\textwidth}{p{1.2cm} >{\raggedright\arraybackslash}X >{\raggedright\arraybackslash}X >{\raggedright\arraybackslash}X}
\toprule
RQ & Primary evidence & Analytic use & Main claims supported \\
\midrule
RQ1 & Coursework and exam instructions, practice rubric, generic marking scheme, planning documents, stored design-time GenAI answers, external review comments & Reconstruct iteration-by-iteration changes to both question authoring and response format, the problems they responded to, and the rationale for tightening the design. & Why the design evolved from generic and open-ended tasks into a more scenario based, constrained, and AI-aware three-part assessment structure. \\
RQ2 & Stabilized rubric, generic marking scheme, sample GenAI answers, worked question materials, documents for markers & Examine how AI-aware question design, source traceability, student revision, and bounded judgement became visible in the response and therefore assessable. & How the design operationalized transparency, reviewability, and evaluative judgement. \\
RQ3 & question level coursework marks, module-result records, design notes, marking observations, external review comments & Describe attainment distributions, note implementation trade-offs, and identify the main workload, fairness, and markability consequences. & Whether the assessment showed a trivial ceiling pattern and what operational costs accompanied the design. \\
\bottomrule
\end{tabularx}
\end{table*}

\subsection{Analytic Procedure and Trustworthiness}
To answer RQ1 and RQ2, we organized the qualitative corpus into an iteration by design goal matrix. For each of the four deployments, we extracted from the archived materials (i) the question or scenario design in use, (ii) the response structure in use, (iii) the operational problems or risks identified, (iv) the design changes introduced, and (v) the rationale or subsequent consequence documented in later artifacts. We then grouped these extracts against the six design goals in Table~\ref{tab:principles}. This produced the refinement account reported in Table~\ref{tab:iterations} and the design claims reported in the Findings section.

To answer RQ3, we analyzed two quantitative views separately. Across the four deployments, the three-part response format appeared in one coursework practice question, two assessed coursework questions, three practice exam quiz questions, and five final exam questions. In the coursework archive, the two assessed coursework questions were stored as six manually graded fields, three fields per question corresponding to the three-part response structure. The first quantitative view therefore uses question level coursework marks for those six graded fields ($n=251$), allowing us to examine the distribution of performance on the SQL and database design question groups. The second view uses official module-result records and complete component profiles ($n=263$ complete cases), allowing us to summarize component- and module-level attainment. The underlying rubric logic for the three-part tasks used a common 18-point internal structure, but those criteria were scaled to the local mark value of each deployed question; in the exam, for example, the same rubric logic was converted to either 6 or 9 marks depending on question weighting. Because the question level coursework analyses use this common rubric logic whereas the module components are reported out of 100 and serve different assessment purposes, we report them separately and do not interpret them as directly comparable effect sizes. We use these distributions only to ask whether marks were trivially compressed at the top end, not to claim superiority to prior years or to assessments that did not use the three part response format.

We added one supplementary response trace analysis from the archived practice response export. After stripping template labels from the three text boxes, one author conducted an initial descriptive coding pass over each attempted \texttt{X1/X2/X3} triad for completion state. For this audit, we used a simple minimum content rule, set during initial inspection of the export, to distinguish substantively attempted fields from blanks, headings, template remnants, or minimal fillers. We then restricted the descriptive coding to the 22 complete triads because only that subset contained all three analytically relevant moves: sourcing, revision, and judgement. Within those complete triads, we noted whether students (i) made an explicit technical critique or correction of \texttt{X1}, (ii) reintroduced lecture or learning-outcome language, (iii) explicitly judged \texttt{X1} broadly sufficient, or (iv) carried source material across fields with little transformation. This response trace work should be read as a lightweight descriptive audit of the archived practice exam quiz sample, not as a standalone qualitative study. The codes are descriptive and non-exclusive; they were used to characterize response patterns and the need for redesign, not to produce a formal qualitative theory or stable prevalence estimates. Appendix~\ref{app:supplement} summarizes the coding cues and provides anonymized exemplar triads.

We also acknowledge a positionality issue. We were closely involved in the design and delivery of the assessment, which gave us direct access to the decision process but also creates interpretive risk. We therefore treat this as reflective practitioner research rather than detached observation. Where possible, we triangulate claims across more than one source type, for example by pairing design notes with archived rubrics, by checking descriptive codes against the worked exemplars and rubric language, or by pairing marking observations with official outcome data and external review comments. Claims based primarily on staff observation are retained only as contextual observations and are discussed cautiously.

\subsection{Ethics, Student Privacy, and Data Governance}
The design required students to submit evidence of tool use, including screenshots of web or GenAI interactions, as part of routine assessment. Those materials were collected first for teaching, marking, and quality assurance purposes rather than for public dissemination, and the present paper reports only a secondary analysis of the archived record. That improves traceability, but it also raises privacy and platform governance questions that need to be stated explicitly. For the present paper, we analyzed those materials only in archived form and reported them either in aggregate or as de-identified, normalized excerpts. We do not reproduce raw screenshots, account details, student names or IDs, or personally identifying interface traces. The appendix exemplars were selected from the archived practice response export from the practice exam quiz, stripped of identifying information, and lightly edited for readability without changing their substantive meaning.

The paper relies on three safeguards. First, only materials already generated through routine module operation were analyzed; no new student facing data collection was added for the case study. Second, the published account uses aggregate summaries and de-identified excerpts rather than public release of raw submissions. Third, the assessment design aimed to avoid making paid access a prerequisite for success: students were not required to purchase a subscription tool, and the exam instructions were framed around tools available through free access. Institution specific governance details are withheld in this anonymized review version because they may identify the site, but the present analysis should be understood as a de-identified secondary use of routine educational records and archived assessment artifacts rather than as publication of raw student trace data.

\section{Findings}
\noindent We present the findings in three layers: what became visible in the response format, what changed across iterative refinement, and what the descriptive outcome data can and cannot say. Throughout this section, we distinguish observed patterns from the broader design implications taken up later in the Discussion. The response trace analysis in particular should be read as a descriptive audit of the archived practice exam quiz sample, not as a standalone qualitative study.

\subsection{The Framework Made AI Use Visible and Assessable}

Addressing RQ2, the clearest observed change was structural rather than statistical: the combination of AI-aware question design and the \Xott{} response structure changed GenAI use from an unrecorded part of students' drafting process, visible to markers only indirectly through the final answer, into part of the assessed performance itself. Students seeking full credit could not rely on a plausible product alone. They had to provide a source trace, show evidence of the interaction, produce a revised answer, and explain why that revised answer was better or why the sourced answer was already sufficient. Taken together, these observations indicate only that, relative to a product only submission format, the framework made students' sourcing, revision, and justification more visible to markers and therefore more assessable.

This interpretation is supported by the archived response traces. In the 22 complete triads from the archived practice response export, 20 contained an explicit technical critique or correction of the sourced answer, 14 reintroduced lecture or learning outcome language, and 5 explicitly judged the sourced answer broadly sufficient rather than rejecting it by default. Table~\ref{tab:workedexample} shows the intended logic of the format, and Table~\ref{tab:appendixexamples} provides anonymized examples of correction, lecture alignment, and justified acceptance.

This mattered especially in database systems because the task family contains several distinct points at which plausibility and correctness diverge. Sourced answers for SQL questions often identified an appropriate join path or grouping idea, but sometimes counted the wrong entity or omitted a discriminator such as term. On modeling tasks, sourced answers could be fluent yet weak at the level of conceptual decomposition. Stronger student responses were distinguishable not because they avoided tools altogether, but because they revised those superficially plausible answers and justified the revision. In that sense, the framework made visible a quality distinction that would have remained hidden in a simple product only submission.

\subsection{Iterative Refinement Was Necessary for Marking Quality and Student Performance}
Addressing RQ1, our second finding is that allowing GenAI is not enough; the assessment format must be engineered carefully if it is to remain educationally useful at scale. In the coursework phase, several operational problems became immediately clear. Students often produced overly long answers, mirroring the verbosity of GenAI systems. Attachments arrived in inconsistent formats, including images, PDFs, and word-processed documents, which increased marking friction. Some students wrote little critical reflection, simply stating that the AI answer was satisfactory and that there was little more to add. Others, unfamiliar with the format, appeared to rush and leave sections incomplete.

The archived practice-response export makes those early problems visible. Twenty-five students submitted responses to the practice exam quiz, which contained three Web/GenAI questions. The maximum possible number of \texttt{X1/X2/X3} triads was therefore 75 (25 students $\times$ 3 questions). The archive, however, contained 50 attempted triads, meaning that not every respondent attempted every question. Here, a triad denotes one student's three-part response to one practice question. Using the completion coding described above, only 22 triads (44.0\%) were substantively complete across \texttt{X1}, \texttt{X2}, and \texttt{X3}; 11 (22.0\%) were effectively \texttt{X1}-only; 8 (16.0\%) reached \texttt{X2} but not \texttt{X3}; and 9 (18.0\%) were partial or minimal. At student level, 11 of the 25 respondents produced at least one complete triad, but only 5 completed all three practice questions with substantive content in all three fields. Attempt counts also declined across the first, second, and third practice questions (24, 15, and 11 attempted triads respectively), suggesting that some students sampled the format without completing the full practice exam quiz.

At the same time, the complete triads showed why the format was worth refining rather than abandoning. In the light descriptive coding of the 22 complete triads, 20 contained an explicit technical critique or correction of the sourced answer, 14 explicitly reintroduced lecture or learning outcome language, 5 judged \texttt{X1} broadly sufficient, and 1 was close to a near carryover of source material across fields. These patterns do not show that students broadly improved as critical thinkers; they do show that, when the full cycle was completed, the assessment captured observable acts of critique, justified acceptance, and technically motivated revision. Table~\ref{tab:appendixexamples} provides three anonymized exemplars of those response moves.

The later refinements directly addressed those problems. The \texttt{X1/X2/X3} structure reduced ambiguity and, in staff judgement, worked particularly well as a stable response template: \texttt{X1} for the AI extract with screenshot evidence, \texttt{X2} for the student's own answer, and \texttt{X3} for narrative and assessment. Word limits curtailed verbosity and forced students to prioritize what mattered. The requirement to rate the sourced answer and justify that rating made non-committal reflection less acceptable. Mandatory full screen screenshots with timestamps improved the evidence trail and simplified marking. A standardized rubric also helped distribute expectations more consistently across markers. In short, the pedagogic idea was viable from the start, but the usable version emerged only after iterative tightening.

At the same time, this productivity gain for assessment validity did not come for free. The design of the tasks, the calibration of the rubric, and the subsequent marking required substantial effort from the teaching team. In practice, some effort was removed from policing hidden AI use, but significant effort was shifted into assessment design, evidence checking, and more interpretive marking of revised answers and narratives.

\subsection{Question Design: Lessons From Testing Draft Tasks With GenAI}
Addressing RQ1 and RQ2, a further finding was that question design itself functioned as a validity screen. Generic or textbook-like questions were often too easy for contemporary GenAI systems to answer at a superficially acceptable level. When a prompt was familiar, decontextualized, or underspecified, GenAI could frequently produce a plausible answer through generic pattern matching with little evidence of real disciplinary discrimination. In that setting, the immediate problem was not simply covert tool use; it was that the question no longer reliably elicited the intended learning outcome.

The module team's question stress-testing clarified this problem. Internal GenAI trials showed that SQL questions could often be answered with a competent looking query skeleton, database design questions could attract plausible but shallow decompositions, and conceptual questions could generate fluent prose that omitted precisely the distinctions the module intended to assess. These observations triggered iterative design. Candidate exam questions were revised when GenAI produced almost complete answers with little effort, when the answer looked persuasive but relied on stock patterns rather than reasoning, or when the task lacked enough contextual specificity to force judgement. The surviving archive supports this pattern across both coursework and exam authoring, even though it does not preserve a complete question by question log of every discarded draft.

The most effective revisions moved the question set toward scenario based, context rich, and module specific prompts. Instead of relying only on generic textbook wording, questions embedded locally authored schemas, short professional scenarios, module specific terminology, and subtle constraints or discriminators that had to be interpreted. This did not prevent GenAI from being useful. Rather, it made direct answer retrieval less reliable and shifted the task toward interpretation, comparison, and judgement. In practical terms, question design and response format played complementary roles: AI-aware question design reduced trivial solvability, while the three-part response structure exposed how students used the tool once the task had been set.

\subsection{Descriptive Outcome Patterns Did Not Show an Obvious Ceiling Effect}
Addressing RQ3, the quantitative evidence is descriptive only. It does not show that the design outperformed prior years or alternative formats, but it does allow one narrower question: did the AI-integrated tasks collapse into a trivial ceiling pattern once GenAI use was built in? Our data suggest that they did not. We report coursework question groups and module components separately because they sit on different scales and answer different questions. In the coursework question level analysis ($n=251$), the AI-supported SQL question group had a mean of 11.50/18 (SD 4.20; median 12), whereas the AI-supported database design question group had a mean of 6.55/18 (SD 4.91; median 6). Only 27.9\% of students scored at least 15/18 on the SQL group, and only 10.0\% reached that level on the database design group. This is consistent with staff observations that stronger students could use sourced material productively, but weaker students still struggled to refine it into technically convincing answers.

A second descriptive view comes from the assessed coursework component itself. Across the two coursework final-grade exports ($n=260$), the normalized MCQ subtotal had a mean of 0.767 (SD 0.129; median 0.792), whereas the AI-integrated essay subtotal had a mean of 0.486 (SD 0.215; median 0.472). This is not a like for like validity comparator, but it does provide an internal contrast within the same coursework: the AI-integrated written component was not trivially maximized relative to the accompanying objective question subtotal. Using half marks on the essay subtotal only as a pragmatic descriptive anchor rather than a validated competence threshold, 48.5\% of students reached at least 0.5, while 17.3\% fell below 0.3. This is broadly consistent with marker judgement that roughly half the cohort could use GenAI, produce their own answer, and sustain a meaningful narrative under the constrained format. The most conceptually demanding database design items were especially discriminating, with zero rates of 53.4\% and 69.7\%, compared with 0.4\% on one of the more procedural SQL items.

The broader module outcomes tell a similar story at component level. The official module result export contained 275 student records; 263 had complete marks across examination, coursework, and practical components and were therefore used for descriptive summary. Within those complete records, the mean mark was 57.52 for the examination component (SD 14.44; median 58.6), 67.29 for the GenAI-enabled coursework component (SD 15.66; median 70.0), and 63.50 overall (SD 12.71; median 65.9). These outcomes are not evidence that the framework alone caused a particular mark distribution; the narrower point is simply that the module did not collapse into uniformly high attainment once GenAI was formally integrated.

Table~\ref{tab:outcomes} summarises these descriptive outcome indicators across the different task scales.

\begin{table}[h!tbp]
\caption{Descriptive outcome indicators reported separately by task scale}
\label{tab:outcomes}
\centering
\footnotesize
\begin{tabularx}{\columnwidth}{@{}p{1.8cm} c c c c >{\raggedright\arraybackslash}X@{}}
\toprule
Measure & Scale & $n$ & Mean (SD) & Median & High-attainment indicator \\
\midrule
AI-supported SQL group & /18 & 251 & 11.50 (4.20) & 12.0 & 27.9\% at 15+/18 \\
AI-supported design group & /18 & 251 & 6.55 (4.91) & 6.0 & 10.0\% at 15+/18 \\
Exam & /100 & 263 & 57.52 (14.44) & 58.6 & 15.2\% at 70+ \\
Coursework & /100 & 263 & 67.29 (15.66) & 70.0 & 51.3\% at 70+ \\
Overall & /100 & 263 & 63.50 (12.71) & 65.9 & 30.0\% at 70+ \\
\bottomrule
\end{tabularx}
\end{table}

\subsection{GenAI Was Most Useful for Structure and Least Reliable for Conceptual Nuance}
The stored GenAI answers revealed a pattern that helps explain why database systems was a particularly revealing site for this design. GenAI was often strong at producing an initial structure, common examples, or a plausible SQL skeleton. It was less reliable when the question required discipline specific framing, conceptual distinctions, or faithful use of module terminology.

For example, on SQL tasks, GenAI often surfaced the appropriate join path and grouping idea, but sometimes failed to count the precise entity required. On database design tasks, GenAI tended to jump quickly to table structures while missing a deeper conceptual issue, such as confusion between conceptual and relational modeling. A particularly revealing pattern was that AI could propose a superficially plausible entity structure such as \texttt{Source(Id, Owner, Provider)} where the better conceptual move was to separate \texttt{Source(Id)}, \texttt{Owner(Id)}, and related entities with explicit relationships. Only a small number of students identified and corrected that level of modelling issue, and those cases were among the clearest examples of the assessment distinguishing genuinely strong performance from merely fluent reuse. On database architecture questions, the AI-generated prose was fluent and persuasive, yet it could still omit key ideas emphasized in the module, such as data independence, tool independence, or the competence relationship between developers and administrators. These examples explain why the ``difference'' and ``assessment'' components of the framework were not decorative extras; they were the core mechanism through which disciplinary understanding was surfaced.

What makes database systems especially analytically useful is that correctness is negotiated across several representational levels. Students must move between scenario interpretation, conceptual modeling, relational structure, and executable SQL. GenAI was often fluent at one level while being weak at another: a query might look syntactically competent while encoding the wrong grouping logic; a design answer might list plausible entities while collapsing distinctions that matter conceptually; an architectural explanation might sound authoritative while omitting the level specific abstractions taught in the module. This multi level character of the domain made it easier to observe the difference between surface adequacy and disciplined reasoning, and therefore made database systems a revealing site for AI-aware assessment design in computing education.

Table~\ref{tab:affordances} summarises the main task-level affordances of GenAI and the recurrent shortcomings surfaced through \texttt{X2} and \texttt{X3}.

\begin{table}[h!tpb]
\caption{Observed task level affordances and recurrent GenAI shortcomings}
\label{tab:affordances}
\centering
\footnotesize
\begin{tabularx}{\columnwidth}{p{1.55cm} >{\raggedright\arraybackslash}X >{\raggedright\arraybackslash}X}
\toprule
Task type & Where GenAI often helped & Recurrent shortcomings surfaced by X2/X3 \\
\midrule
SQL & Join paths, grouping skeletons, and initial query structure & Wrong entity counted, missing discriminators such as term, or incomplete \texttt{HAVING}/aggregation logic \\
DB design & Candidate entities, attributes, and broad decomposition ideas & Conflated conceptual and relational design, bundled entities improperly, or missed normalization and dependency issues \\
Architecture / concepts & Fluent summaries and broad lists of points & Vague disciplinary framing, missing module specific distinctions, and persuasive but incomplete prose \\
\bottomrule
\end{tabularx}
\end{table}

The particularly weak performance on conceptual database design tasks is consistent with recent database education evidence. \citet{jamil2025} reports that LLMs used as database design tutors remain prone to serious reasoning errors on functional dependencies and normalization, while \citet{rizzi2025} show that LLM performance in conceptual data design is uneven and depends heavily on detailed procedural prompting. More importantly for our purposes, these patterns provided the module team with a practical authoring signal: the questions that remained educationally useful were typically those where scenario context, embedded assumptions, and module specific discriminators prevented the AI answer from functioning as a fully adequate substitute for student reasoning.

\subsection{Operational Observations and Boundary Conditions}
Several additional observations are worth recording for educators. First, time pressure changed how students used GenAI. Under tighter exam conditions, and especially where the question stem itself required contextual interpretation, stronger students often chose to write in their own words rather than spend excessive time iterating prompts in an attempt to force GenAI into the required format. This is relevant because it suggests that the combined design can change the local economics of tool use: beyond a certain point, continued prompting becomes less efficient than making a disciplined judgement directly.

Second, there was a wide range in the quality of the sourced material that students selected for \texttt{X1}. Importantly, the marking of \texttt{X1} did not reward the objective quality of the AI system's answer as such; rather, it rewarded whether the student extracted something relevant and documented it properly. This distinction matters because the assessment was intended to evaluate student judgement, not to rank AI systems.

Third, a non trivial number of students appeared to stop after \texttt{X1}, providing an extract and evidence but leaving \texttt{X2} and \texttt{X3} absent or very weak. Relatedly, some students seemed to position the AI answer as superior to anything they themselves could produce, which worked against the evaluative purpose of the task. Finally, fairness required attention to access conditions: the exam design explicitly avoided making paid GenAI subscriptions an advantage, aligning with the sector concern that some tools may sit behind paywalls \cite{russell2023}. Screenshot evidence also improved reviewability of the response process during marking, but we treat that only as a minor operational benefit rather than as evidence about wider integrity effects.

\section{Discussion}
\subsection{Assessment Design Contribution}
This paper contributes a reusable assessment design for computing education. The design combines AI-aware question authoring before deployment with a three-part response structure (\texttt{\Xott}) during assessment completion. Unlike broader policy oriented calls for adaptation \cite{chan2023,luo2024,xia2024}, the contribution here is operational: it specifies how educators can test whether draft questions are too easily completed by current GenAI systems, and how students' use, revision, and evaluation of sourced material can then be made visible in the assessed response.

The response structure contribution lies in requiring students to document, revise, and evaluate sourced material as part of the assessed response. The \texttt{\Xott} structure does not reward students simply for retrieving fluent text; it requires a chain of actions that can be marked separately: sourcing, evidencing, revising, and judging. We do not claim that this structure measures broad development in critical thinking. Rather, its contribution is to make evaluative judgement and technically grounded revision more visible within the submission than would be possible in a product only format.

The question authoring contribution addresses a different part of the same validity problem. During authoring, we used contemporary GenAI tools as stress testing baseline: draft the task, attempt it with available tools, inspect where the answer is deceptively plausible, revise the task when superficial prompting is too effective, and retest before deployment. The most effective revisions moved questions toward scenario based, context rich, and module specific formulations with embedded constraints and discriminators. This is not a claim that assessment can be made immune to AI. Rather, it is a practical way of making questions less susceptible to superficial GenAI completion and better aligned with intended learning outcomes.

The case study then adds implementation knowledge about what it took to make that combined design work at scale. Standardized evidence formats, explicit word limits, practice opportunities, reusable marking logic, and common expectations about free tier tool access were not peripheral details; they were part of the method. The practical lesson is that apparent productivity gains do not remove staff effort so much as redistribute it from hidden authorship policing toward question refinement, rubric engineering, and more interpretive marking. In that sense, the paper contributes a design method plus an account of its operational costs and constraints.

The combined design also complements, rather than duplicates, broader AI-era assessment frameworks. AIAS and related sector guidance help educators decide what kinds of AI use should be permitted, signposted, or redesigned \cite{perkins2024,qaa2023reconsidering}. Our contribution is a more granular answer to two linked questions: once AI use is explicitly allowed in a technical assessment, what response structure makes that use visible and markable, and what authoring process helps educators design questions that remain educationally productive under current AI capability?

\subsection{Interpreting the Findings Through Existing Literature}
The findings align well with authentic assessment literature, but they also extend it. \citet{villarroel2018} and \citet{sotiriadou2020} emphasize the value of authenticity for meaningful learning and employability. Our results support that position: letting students use web and GenAI tools made the tasks feel closer to real problem solving in contemporary technical settings. However, the findings also support the warning from \citet{ellis2020} that authenticity alone is insufficient for academic integrity. What mattered here was not merely that the assessment was authentic, but that authenticity was paired with process evidence, explicit explanation of differences, and a forced judgement of quality.

We also extend the literature on digital authenticity \cite{nieminen2023,ajjawi2024}. In a GenAI context, authenticity is not only about whether the task resembles future work. It is also about whether the student must make consequential judgements within that digitally mediated environment, and whether the question itself contains enough contextual richness to require that judgement. A task that merely permits GenAI may be authentic in one limited sense, but if it does not require students to examine the quality and limitations of what they retrieve, it misses a central dimension of authentic professional practice. Seen through constructive alignment, this is crucial: the assessment only remains valid if both the question and the response format elicit the intended learning outcomes \cite{biggs1996}.

This is where the AI-aware question design contribution matters theoretically. Much of the current literature correctly recommends more authentic and contextualized assessment, but our case shows a practical mechanism for enacting that advice: use GenAI during question authoring as a stress testing baseline, identify where draft tasks are vulnerable to superficial completion, and revise them toward scenario based, context rich formulations. The contribution is therefore methodological as well as pedagogical. It offers an operational bridge between general calls for redesign and day to day assessment authoring practice.

Finally, the findings reinforce \citet{tai2018}'s argument that evaluative judgement is a crucial educational capability. The rating and rationale element in \texttt{X3} was not just a reflective flourish; it compelled students to decide whether the sourced answer was excellent, adequate, or weak, and to justify that assessment briefly. In GenAI era, that capacity may be at least as important as the ability to draft an answer from scratch. We treat this as an interpretation of what the assessment made visible, not as proof that students developed a broader disposition through the intervention alone. In that narrower sense, the findings are also consistent with \citet{facione1990}'s definition of critical thinking as purposeful, self-regulatory judgement, with \citet{zimmerman2002}'s view of learners as active agents who monitor and evaluate their own performance, and with \citet{bearman2024}'s argument that evaluative judgement now includes judging the provenance, prompting, and trustworthiness of AI-assisted work.

\subsection{Implications for Computer Science Educators}
For computer science educators, the key implication is that GenAI friendly assessment needs to be both discipline shaped and AI-aware at the point of authoring. The most effective questions in this module were not generic essay prompts. They were technical questions on SQL, database design, architecture, and data related topics where GenAI could help, but where full marks still required students to notice specific distinctions, constraints, and trade-offs. Database systems is instructive here because it contains both procedural and conceptual task types, allowing educators to see clearly where fluent AI output diverges from disciplined reasoning. In practice, this means that draft questions should be tested against current GenAI systems before deployment, especially when they are scenario based or technically subtle. This is closely aligned with benchmark guidance, which emphasizes critical reflection, communication, professional integrity, and the ability to locate, retrieve, and attribute relevant ideas as part of disciplinary competence rather than as peripheral skills \cite{qaa2022computing}.

The second implication is that process must be assessed explicitly. If educators want students to use GenAI critically, then prompt choice, source specification, evidence of interaction, comparison against one's own answer, and justification of quality all need to be visible parts of the response. Otherwise, the educationally important part of the work remains hidden.

The third implication is operational. Large cohort sustainability depends on standardization. Evidence formats, word limits, and reusable marking structures are not merely administrative conveniences; they are part of how the pedagogy works. Without them, the assessment becomes harder to mark consistently and easier for students to treat as an invitation to submit large amounts of low value AI-generated text. That operational point is especially important because several of the strongest findings in this case study emerged only after the design was tightened around bounded fields, mandatory evidence, and common rubric expectations.

\subsection{Implications Beyond Computer Science}
Although this case comes from database systems, the broader design problem is not discipline specific. Many educators across higher education are struggling with the same tension: students need to become capable, ethical users of GenAI, yet institutions also need credible evidence of individual achievement. The three-part response structure offers one answer to that tension by relocating the focus of assessment from hidden tool use to accountable tool use.

Within universities, such an approach can help align assessment redesign with emerging AI-literacy agendas and fairness concerns, especially where subscription based tools may otherwise create inequitable advantages \cite{russell2023}. It also offers a practical alternative to detector led integrity approaches that remain contested on both accuracy and fairness grounds \cite{liang2023,webb2025detect}. Internationally, the framework is portable because it is not tied to one vendor, one model, or one discipline; its central idea is to assess the student's judgement about sourced material rather than the brand of tool consulted. The same is true of the question design methodology: using GenAI as an stress testing baseline during assessment authoring is a transferable practice for any discipline where generic prompts have become too easy to complete superficially.

\section{Limitations}
Our research has several limitations.

\textit{Construct validity:} we interpret \texttt{X2} revision, explanation of differences, and \texttt{X3} judgement as evidence that the design elicited evaluative judgement and critical review. However, we did not administer an independent instrument for critical thinking or evaluative judgement. Our claims are therefore about what the assessment was designed to elicit and make visible, not about measured gains in those capabilities.

\textit{Corpus and causal attribution:} we report a single module case without a randomized or quasi-experimental comparison. The evidence base is heterogeneous and more complete for some phases than others, especially the later exam iterations. Multiple design changes were introduced across the four iterations, so we cannot isolate the effect of any one change or claim that the framework alone caused the observed mark distributions. Nor can we use these data to show superiority over prior years, over assessments that did not use the three part response format, or over traditional online assessments.

\textit{Cohort characterization:} although routine admissions data indicated that the module was taught in English to a culturally and ethnically diverse undergraduate cohort with substantial representation of both home and international students, we did not conduct a formal demographic analysis. We therefore cannot make subgroup claims about how the design functioned across ethnicity, domicile status, or different language backgrounds.

\textit{Reliability and researcher role:} we were closely involved in designing and delivering the assessment. That is a strength for reconstructing the design process, but it also risks over interpreting success. Some observations, such as the perception that stronger students preferred to write in their own words under time pressure, were not collected through a formal qualitative protocol. The supplementary coding of practice responses strengthens the evidence base, but it remains a single analyst, light descriptive analysis of a limited practice exam quiz sample rather than a full qualitative study. Future work should combine assessment design reporting with more systematic interview, survey, or trace data methods.

We also did not compute a formal inter-rater reliability statistic across markers. The shared rubric and generic marking scheme were designed to support consistency, but we cannot claim reliability evidence beyond that procedural standardization.

\textit{Ethics and data governance:} the design required students to submit screenshots of web or GenAI interactions as assessed evidence. That improves transparency, but it also raises privacy and platform governance issues. For this paper, we report only aggregate outcomes and de-identified excerpts; we do not reproduce raw screenshots or account identifiable traces. Institution specific approval and governance details are withheld in this anonymized review version. The ethics discussion in this paper should therefore be read as a description of the safeguards applied to the present analysis, not as a full normative solution to the broader governance issues raised by external AI tools in assessment.

\textit{Transferability and sustainability:} we report one institutional case in one computing module. The framework is likely adaptable, but transferability should not be confused with proof of universal effectiveness. A further limitation is workload: while the three-part response structure improved visibility and fairness, it required substantial staff effort in both design and marking. The same is true of AI-aware question design: stress testing of draft questions is time consuming, and its usefulness is historically contingent because GenAI systems and interfaces change rapidly. Any implementation will therefore require periodic re-testing and review.

\section{Conclusion}
In this paper, we have described an 
assessment design that makes the usage of AI \emph{mandatory}, instead of viewing the usage of AI as academic misconduct (an exam offence).
This approach is suitable for the GenAI era since it retains
technical specificity and visible grounds for academic judgement.

In a large undergraduate database systems module, previous years had shown increasing problems (collusion and other misconduct) in non-invigilated online exams. 
In an iterative process, we have developed a three-part response format (\texttt{\Xott}) that assessed not just what answer students produced, but how they used, evidenced, revised, and judged sourced web/GenAI material. 

Across several iterations, we moved from a promising pedagogic idea to a more operationally, stable format through tighter evidence rules, clearer structure, explicit judgement tasks, and practical rehearsal opportunities.
Most importantly, there was a process to introduce students to the exam format and rubric, laying out the academic integrity rules.

Our central lesson is that GenAI introduces a layered assessment challenge involving task design, acceptable tool use, evidence of process, marking practice, and questions of fairness and governance. %

Based on what we reported here, we can claim that the combined design of AI-aware question authoring and the \Xott{} format made tool use visible, reviewable in marking, and tightly connected to evaluative judgement.

We do not claim that we have directly measured gains in critical thinking; rather, we argue that the framework offers a practical way to make evaluative judgement and revision inspectable in an AI-aware environment. This means setting technically meaningful questions, demanding visible evidence of process, and rewarding judgement rather than surface fluency alone.

In the GenAI era, suitable assessment depends not only on how students are asked to respond, but also on how educators/lecturers design and pre-test the questions themselves against current AI capability.

Overall, the \Xott{} assessment format introduced in this paper is a generic and reusable template to test the AI literacy (X1), the learning outcomes (X2), and the reflective capability (X3) of students in the GenAI era.

\clearpage
\begin{appendices}
\renewcommand{\theHtable}{appendix.\Alph{section}.\arabic{table}}
\renewcommand{\theHfigure}{appendix.\Alph{section}.\arabic{figure}}
\section{Supplementary Coding Protocol and Anonymized Exemplars}
\label{app:supplement}

Table~\ref{tab:codebook} records the lightweight coding rules used for the archived practice-response analysis. The protocol was intentionally simple because the goal was descriptive auditability rather than formal grounded-theory development. Completion-state coding relied on a pragmatic minimum-content screen established during initial inspection of the export so that near-empty fields would not be treated as substantive responses. Student identifiers were removed throughout, and the excerpts in Table~\ref{tab:appendixexamples} were lightly trimmed and normalized for readability without changing their substantive meaning.

\begin{table}[h!tbp]
\caption{Lightweight coding protocol used for the practice-response analysis}
\label{tab:codebook}
\centering
\scriptsize
\begin{tabularx}{\columnwidth}{@{}p{1.55cm} >{\raggedright\arraybackslash}X >{\raggedright\arraybackslash}X@{}}
\toprule
Code & Operational cue & Analytic use \\
\midrule
Full triad & \texttt{X1}, \texttt{X2}, and \texttt{X3} each contain substantive text after template stripping rather than only headings, template remnants, or brief fillers. & Indicates that the student completed the retrieval, revision, and judgement cycle. \\
X1-only / X1+X2-only & \texttt{X1} is substantive but later fields are absent or minimal. & Indicates retrieval without full revision or evaluative follow-through. \\
Technical critique & \texttt{X2} or \texttt{X3} explicitly identifies a defect, omission, or stronger disciplinary alternative to \texttt{X1}. & Evidence that the student moved beyond acceptance of the sourced answer. \\
Lecture alignment & \texttt{X2} or \texttt{X3} reintroduces lecture, learning-outcome, or module terminology absent from \texttt{X1}. & Evidence that the student is reconnecting the sourced answer to course concepts. \\
Endorses X1 & \texttt{X2} or \texttt{X3} explicitly states that little or no change to \texttt{X1} was needed and briefly justifies that judgement. & Boundary case showing that the framework allows justified acceptance rather than compulsory disagreement. \\
Carryover & Source material is repeated across fields with minimal transformation. & Marks the upper boundary of weak evaluative engagement even when the triad is formally complete. \\
\bottomrule
\end{tabularx}
\end{table}

\begin{table}[!hb]
\caption{Anonymized exemplar \texttt{X1/X2/X3} traces from the archived practice-response export}
\label{tab:appendixexamples}
\centering
\scriptsize
\begin{tabularx}{\textwidth}{@{}p{1.2cm} p{1.0cm} >{\raggedright\arraybackslash}X >{\raggedright\arraybackslash}X >{\raggedright\arraybackslash}X@{}}
\toprule
Student & Task & X1 excerpt & X2/X3 excerpt & Why this matters analytically \\
\midrule
A & SQL & Sourced answer used a self-join strategy to find students enrolled in multiple sections with the same instructor in the same term. & Student rewrote the solution using \texttt{GROUP BY} and \texttt{HAVING COUNT(DISTINCT ...)} and explained in \texttt{X3} that ``two or more'' is easier to reason about through grouping than through pairwise self-joins. & Illustrates technical critique and principled revision rather than simple paraphrase. \\
B & DB design & Sourced answer fixed the postcode--city dependency and proposed moving stock information, but did not justify the design with normalization terminology. & Student argued that the answer still missed 1NF/2NF or 3NF issues, introduced functional-dependency language, and explained in \texttt{X3} that the revised answer was better because it justified the change rather than merely naming it. & Illustrates lecture alignment and stronger conceptual reasoning on a harder design task. \\
C & SQL & Sourced answer already used the correct grouped SQL pattern and documented the source clearly. & Student kept the same core query, stated that no major change was necessary, and rated the sourced answer highly while still noting that the grouped logic was what made it correct. & Illustrates a justified endorsement case: the format can capture acceptance with rationale, not only correction. \\
\bottomrule
\end{tabularx}
\end{table}

\end{appendices}
\clearpage

\section*{Declarations}

\subsection*{Ethics approval and consent to participate}
This paper reports a secondary analysis of routine educational records and archived assessment artifacts generated during normal module delivery, marking, and quality-assurance processes. The study did not recruit participants separately for research purposes and did not introduce an intervention beyond standard teaching practice. The paper reports only aggregate results and de-identified excerpts, rather than reproducing raw screenshots or account-level traces. On that basis, formal ethics approval was not required for this paper under local institutional arrangements governing routine educational evaluation and secondary analysis of de-identified teaching records.

\subsection*{Consent for publication}
Not applicable. No identifiable personal data, raw screenshots, or account-level traces are reproduced.

\subsection*{Availability of data and material}
The student-level assessment records, screenshots, and internal module documents analyzed in this study are not publicly shared because they form part of live educational provision and contain potentially identifying student data, interface traces, and assessment-security information. The paper reports aggregate outcomes and de-identified excerpts only. Anonymized supporting materials sufficient to understand the assessment design can be made available to the editor on reasonable request, subject to institutional constraints.

\subsection*{Competing interests}
The authors declare no competing interests.

\subsection*{Funding}
The authors received no specific funding for this work.

\subsection*{Authors' contributions}
Riasat Islam led the assessment redesign, curated the documentary and assessment evidence, conducted the descriptive analysis, and drafted the manuscript. Thomas Roelleke contributed to the assessment design, conceptual framing, interpretation of the evidence, and manuscript revision. Both authors approved the submitted version.

\subsection*{Acknowledgements}
The authors thank colleagues, Frederik Dahlqvist and Syed Rafee, who were involved in module delivery, marking, and review for feedback that informed the iterative refinement of the assessment design reported in this paper.

\clearpage
\bibliographystyle{apalike}
\bibliography{references}

\end{document}